\documentclass[a4paper,10pt]{article}
\usepackage{xcolor}
\usepackage{amsmath}
\usepackage{hyperref}
\usepackage{cleveref}
\usepackage{amsfonts}
\usepackage{bbold}
\usepackage{graphicx}
\usepackage{geometry}
\newcommand{\DNF}{\mathcal{D}^{\rm NF}}
\newcommand{\pnf}{p^{\rm NF}}

\newcommand{\DMLP}{\mathcal{D}^{\rm MLP}}
\newcommand{\Umlp}{U^{\rm MLP}}

\newcommand{\xyz}{x}
\newcommand{\ic}{z}
\newcommand{\rc}{y} 
\newcommand{\R}{\mathbb{R}}
\newcommand{\FtoIC}{f}
\newcommand{\state}{a}

\newcommand{\Edft}{U^{\rm DFT}}
\newcommand{\Eneutral}{U}
\newcommand{\Ebase}{U_{\rm B}}
\newcommand{\Zbase}{Z_{\rm B}}
\newcommand{\pbase}{p_{\rm B}}

\newcommand{\Zdft}{Z}

\author{Ana Molina-Taborda$^{1,2,4}$ \and Pilar Cossio$^{3,4,*}$ \and Olga Lopez-Acevedo$^{1,2,*}$ \and Marylou Gabrié$^{5,*}$}
\date{ \small
$^1$ Biophysics of Tropical Diseases Max Planck Tandem Group, University of Antioquia UdeA, 050010  Medellin, Colombia \\
$^2$ Grupo de Física Atómica y Molecular, Facultad de Ciencias Exactas y Naturales, Universidad de Antioquia UdeA, 050010  Medellin, Colombia \\
$^3$ Center for Computational Mathematics, Flatiron Institute, New York, USA \\
$^4$ Center for Computational Biology, Flatiron Institute, New York, USA\\
$^5$ CMAP, CNRS, École polytechnique, Institut Polytechnique de Paris, 91120 Palaiseau, France\\
*pcossio@flatironinstitute.org, *olga.lopeza@udea.edu.co, *marylou.gabrie@polytechnique.edu
}
\title
{Active learning of Boltzmann samplers and potential energies with quantum mechanical accuracy}

\begin{document}
\maketitle



\begin{abstract}
Extracting consistent statistics between relevant free-energy minima of a molecular system is essential for physics, chemistry and biology. Molecular dynamics (MD) simulations can aid in this task but are computationally expensive, especially for systems that require quantum accuracy. To overcome this challenge, we develop an approach combining enhanced sampling with deep generative models and active learning of a machine learning potential (MLP). We introduce an adaptive Markov chain Monte Carlo framework that enables the training of one Normalizing Flow (NF) and one MLP per state, achieving rapid convergence towards the Boltzmann distribution. Leveraging the trained NF and MLP models, we compute thermodynamic observables such as free-energy differences or optical spectra. We apply this method to study the isomerization of an ultrasmall silver nanocluster, belonging to a set of systems with diverse applications in the fields of medicine and catalysis.
\end{abstract}



\section{Introduction}

The understanding of a wide variety of phenomena such as chemical reactions, nanodevice design, or protein folding builds on extracting an accurate and consistent 
atomistic thermodynamic description of the system  \cite{Aspuru-Guzik_Ceriotti_2024}. These systems typically interconvert between several metastable states (relevant to function under intended operating conditions) that are separated by a large free-energy barrier, and that can be activated upon changes in the environment or interactions with other molecules. For example, quantifying the interaction free-energy between biomolecules and nanoclusters can aid the determination of the nanoclusters' uptake, toxicity and detoxification mechanisms when used for antibiotic treatment, for imaging and magnetoresponsive therapy and for cancer therapy \cite{Frei23, Lee15, Wang2022, Weilian18}. Experimental measurements, such as optical spectra, typically provide bulk ensemble averages, where disentangling the contribution of each metastable state and determining the underlying molecular mechanisms is challenging.   

In principle, atomistic and mechanistic insights could be obtained using molecular dynamics (MD). However, the states associated with the free-energy minima are typically separated by a high free-energy barrier, which is difficult to overcome with simulations based on thermal fluctuations. For barriers larger than
$10 k_BT$, the residence time in each state is orders of magnitude larger than the total possible time the system can be simulated for with available computational resources. The problem becomes even more difficult when studying systems using \textit{ab initio} quantum accuracy, where the dimensionality of the system's representation significantly increases. 
These issues render the sampling of multi-state systems, along with the corresponding free-energy calculations, extremely challenging at \textit{ab initio} accuracy \cite{ceriottiPotentialsIntegratedMachine2022}.

With \textit{ab initio} methods, the free-energy differences between two isomers are typically estimated using an ideal-gas thermodynamic approximation \cite{cramer2004}. This approach consists in calculating the free-energy difference using the energy and vibrations of an optimized structure in the zero-temperature ground state. This method may not be always valid because it approximates the vibrational modes that contribute to the entropy and internal energy as quantum harmonic oscillators. To understand systems at a quantum level description, enhanced sampling methods that explore conformational spaces efficiently have also been applied  \cite{biarnes2007,piccini2022ab,schilling2020zooming,yuan2023accurate,devergne2022combining}. For example, \textit{ab initio} metadynamics was used to explore the conformational landscape of silver nanoclusters and extract the free-energy surface along a small set of collective variables (CVs) \cite{sucerquia2022}. However, these methods are still computationally expensive and require the choice \textit{a priori} of the biasing CVs. 

On the other hand, machine-learning (ML) has revolutionized many scientific fields, including quantum chemistry. 
Machine-learning potentials (MLPs) are learned force fields with quantum accuracy that are widely used in MD simulations \cite{behlerFourGenerationsHighDimensional2021,zubatiuk2021development,xie2023ultra,schran2021machine,ko2021fourth,behler2021machine}. They are also starting to be utilized to compute thermodynamical quantities \cite{ceriottiPotentialsIntegratedMachine2022,chen2023constructing,chengThermodynamicsDiamondFormation2023}, for example, using Markov chain Monte Carlo (MCMC) for surface reconstruction \cite{du2023machine}.
ML approaches have aided conformational sampling by learning CVs for biasing MD simulations \cite{bonati2021deep,ribeiro2018reweighted,wang2020machine}, learning dynamics \cite{chen2023constructing}, committors \cite{rotskoffActiveImportanceSampling2022, jung2023machine} or by elucidating transition states \cite{duan2023accurate}.
Generative ML models have also aided conformational sampling, with the seminal Boltzmann generators proposing -- along with other pioneering works in statistical mechanics \cite{albergoFlowbasedGenerativeModels2019,wuSolvingStatisticalMechanics2019} -- to emulate the Boltzmann distributions of molecular systems with Normalizing Flows (NFs) \cite{noe_boltzmann_2019}. 
After learning, the NF model is typically combined with Monte Carlo strategies to guarantee the accuracy of estimators built with flow samples. Among the possible approaches, the adaptive MCMC assisted with NFs \cite{gabrieAdaptiveMonteCarlo2022} -- sometimes referred to as flowMC -- is remarkably efficient in terms of the number of energy calls, since training and sampling are performed simultaneously. 

NFs are also known to greatly aid in free-energy computations. Several works advocate training NFs on MD trajectories to learn mappings between states of interest
and a reference \cite{wirnsberger_targeted_2020, ding_computing_2020, wirnsbergerNormalizingFlowsAtomic2022}.
The learned mappings are then leveraged to perform targeted free-energy perturbation (TFEP) \cite{jarzynski_targeted_2002,wirnsberger_targeted_2020} or Bennett acceptance ratio (BAR) \cite{Bennett1976,jia_normalizing_2019,ding_computing_2020} calculations yielding estimates of the free-energy differences. 
However, these methods require preexisting and sufficiently well-sampled (at least around the modes) MD trajectories for training the NF, where sometimes the MD trajectories themselves follow less accurately the desired Boltzmann statistics. Moreover, these methods have not yet been applied for \textit{ab initio} simulations, which are inherently much more time-consuming than classical ones. 

To overcome these limitations, we develop a sampling, potential learning, and free-energy calculation framework for \textit{ab initio} simulations which builds on flowMC \cite{gabrieAdaptiveMonteCarlo2022} by adding the simultaneous active learning of an MLP. The main ingredients of the strategy, coined ab-flowMC, are NFs to propose configurations around free-energy minima, MLPs to approximate DFT energies, and MCMC to access the Boltzmann distribution. We validate the method, demonstrating its gain in accuracy and computational efficiency to study the thermodynamic 
properties of a silver nanocluster due to the wide potential of such materials in technological applications \cite{Frei23, Wang23, Anzhela21, Liu23, Morones-Ramirez2013, zhangEngineeringHoleTransporting2023, chrysouliWatersolubleSilverFormulation2018}. 

\section{Results}

\subsection{The ab-flowMC method}

The \textit{ab initio} flowMC (ab-flowMC) method aims at producing Boltzmann distributed configurations across several free-energy minima of a molecular system, such as isomers. For simplicity, we assume that there are only two states of interest, but the framework can be generalized to multiple states. We also assume that one initial configuration of the molecule in each state is known (for instance a local energy minimum), such that short MD simulations can first be produced. Ab-flowMC then consists of training locally for each state of interest an MLP and an NF for approximating the \emph{ab initio} energies and the Boltzmann distribution, respectively. 
Although we could train one NF and one MLP across all states, the corresponding functions to learn are more complex than when training one pair of models per state, which therefore requires more training samples. As such, we learn different models for each state to reduce the overall number of \emph{ab initio} energy calls that are required to build the training data -- since they dominate the computational budget--.
Moreover, using one NF per state also stabilizes training in cases where one state is much rarer than the other, and therefore, difficult to capture in a joint learning procedure.

\begin{figure}[t]
\centering
\includegraphics[width=\textwidth]{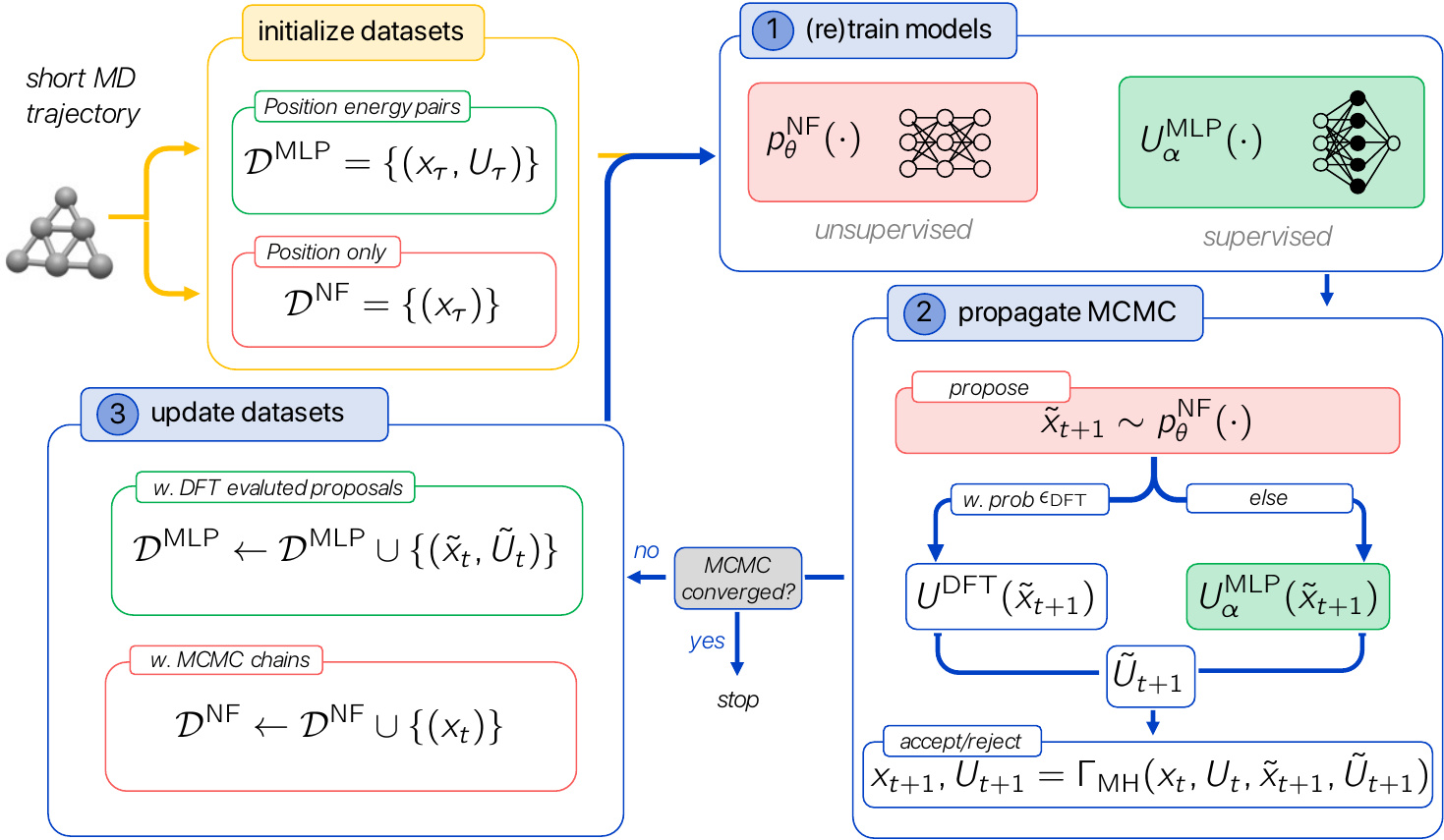}
\caption{\emph{Ab initio} flowMC workflow. 
Starting from short MD trajectories, a cycle of three steps is repeated. Step 1, the normalizing flow (NF) and the machine learning potential (MLP) are trained. Step 2, leveraging the NF and MLP models, MCMC chains are generated by restarting them from the last state of the previous cycle. Step 3, the training datasets of the ML models are augmented with the newly created MCMC samples and the position-energy pairs for which DFT was evaluated.}
\label{fig:algo-sketch}
\end{figure}

We depict the ab-flowMC workflow in \cref{fig:algo-sketch} for a single state. 
The algorithm starts with a short MD trajectory at the temperature of interest using the desired level of accuracy (henceforth, we consider DFT for simplicity). The generated positions along the trajectory, indexed by $\tau$, are collected to initialize a dataset $\DNF = \{(x_\tau)\}$, while the position-energies pairs are collected to initialize a second dataset $\DMLP = \{(x_\tau, U^{\rm DFT}_\tau)\}$.
We denote by $\pnf_\theta$ the probability density defined by the NF on the configuration space, and $\Umlp_\alpha$ the MLP defined by a feed-forward neural network. The positions of the atoms are defined using internal coordinates \cite{chemcoord-paper} to directly account for translation and rotational invariances. Note that ML models with architectures designed to respect these invariances could also be used to work directly with Cartesian coordinates. 

A cycle of three steps is repeated until either an MCMC convergence criteria is met or the ML models are sufficiently trained for subsequent applications. First, the dataset $\DNF$ is used to train the NF parameters $\theta$  by maximum likelihood, and the dataset $\DMLP$ is used to train by least squares regression the MLP parameters $\alpha$. Second, an MCMC simulation is propagated for several steps. At each step $t$, the NF is sampled to propose a new configuration $\tilde{x}_{t+1}$.
The energy computation method of $\tilde{x}_{t+1}$ is selected at random between DFT, with probability $\epsilon_{\rm DFT}$, or approximation with the MLP, with probability $1-\epsilon_{\rm DFT}$. The Metropolis-Hastings criteria is used to accept or reject the configuration (see the Methods). To boost exploration, multiple MCMC chains are run in parallel (indices not denoted). Third, the dataset $\DNF$ is extended with the newly produced MCMC configurations while the dataset $\DMLP$ is augmented with the proposed position-energy pairs for which DFT was evaluated, even if they were rejected\footnote{Note that it is important to also keep the rejected pairs to ensure that the MLP is accurate enough to trigger rejections in cases where the NF proposes high-energy configurations.}. The cycle then repeats.
At each cycle, the MCMC chains are restarted at the last configurations of the previous cycle. As such, ab-flowMC belongs to the class of adaptive MCMC procedures: 
NF training along the iterations improves the Metropolis-Hastings Markov kernel. Indeed, provided that the MLP predicts reliably the potential energies, the stationary distribution of the ab-flowMC Markov chains is the Boltzmann distribution, which ensures that the flow training data is following, more and more closely, the Boltzmann distribution. Ab-flowMC also incorporates active learning of the MLP, as commonly employed in the \textit{ab initio} chemistry community, although here the computation of new training points is dictated by the flow proposals instead of MD simulations. 
The MLP accuracy is closely related to the amount of data used for training. In this regard, a compromise between computational costs and convergence is found by decreasing the value of $\epsilon_{\rm DFT}$ along the cycles.
As always in ML applications, tracking overfitting is crucial; our monitoring strategy is detailed in the Supplementary Text \ref{app:sec:overfitting}.

The ab-flowMC actively-learned Boltzmann-like distributions and potential energies can be used in rich and diverse ways to study molecular systems with \textit{ab initio} accuracy. The advantages lie in having NFs that produce candidate samples efficiently with fast potential energy computations using the trained MLPs. We can use them, for example, to compute optical spectra or absolute and relative free energies, or to construct a mixture model of flows (ab-flowMM in the Methods) for sampling mixed isomer states and  extract isomers relative populations. 
 
\subsection{Ab-flowMC study of $Ag_6$}

Despite the commercial use of silver nanomaterials and their wide range of potential applications, from combating antimicrobial resistance, drug carriers, to photocatalytic hydrogen evolution \cite{Wang23, Anzhela21, Liu23, Morones-Ramirez2013,Frei23, zhangEngineeringHoleTransporting2023, chrysouliWatersolubleSilverFormulation2018}, the quantification of silver nanoclusters' thermodynamic properties and molecular mechanisms remains elusive. 
In the following, we demonstrate ab-flowMC's potential to study ultrasmall silver clusters isomerization, a type of problem with nanotechnological relevance  \cite{chrysouliWatersolubleSilverFormulation2018,zhangEngineeringHoleTransporting2023}, using as benchmark a silver cluster that was previously studied with \textit{ab initio} metadyanmics \cite{sucerquia2022}.

\subsubsection{$Ag_6$ isomers and MD simulations}

The experimental optical absorption spectrum of small silver clusters reveals the concurrent presence of multiple isomers at low temperatures, potentially initiating from a nascent stage, specifically at $N_a=6$ atoms \cite{Lecoultre2011}. At $N_a=6$ the lowest-energy isomers that are possibly concurrent are one planar and one bipyramidal non-planar (\cref{fig:fig2}A and Supplementary \cref{app:fig:single-isomer-planar}A, respectively), these configurations have also been previously reported in theoretical studies \cite{Fournier2001, Duanmu2015}. On the other hand, population percentages estimated from computational methods are not in agreement. At 300K, gas limit thermochemistry gives 97\%-3\% assuming only those two isomers are present, while a recent \textit{ab initio} metadynamics predicts this ratio at 90\%-10\%  \cite{cramer2004, sucerquia2022}. 
We chose to study this $Ag_6$ molecule at 350K considering these two main metastable states, the planar and the bipyramidal (denoted isomer 0 and isomer 1, respectively). We first ran \textit{ab initio} MD simulations for 0.007 ns for each isomer (see the Methods). We note that with the available computational resources, it is not possible to transition between these metastable states using pure MD. 

\subsubsection{Ab-flowMC for each $Ag_6$ isomer}

\begin{figure}[h]
\centering
\includegraphics[width=\textwidth]{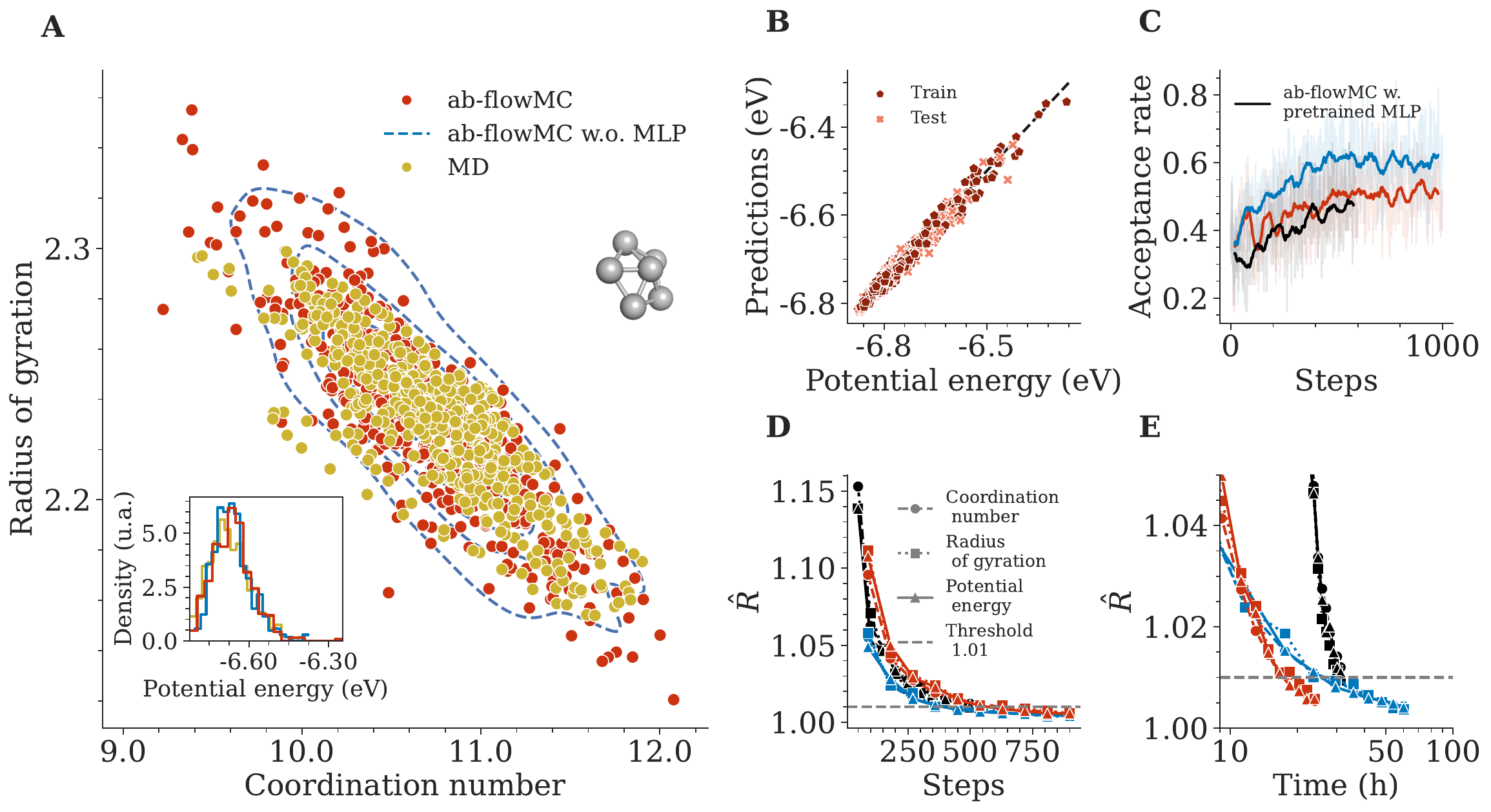}
\caption{\textit{Ab initio} flowMC for the $Ag_6$ bipyramidal isomer 0. A) Projection along the coordination number and radius of gyration of samples from the MD trajectory (yellow) and the ab-flowMC simulation (red). The density of samples obtained from the ab-flowMC w.o. MLP simulation is displayed for reference (blue dashed lines). Inset: Histograms of the potential energies for all methods. B) DFT potential energy (x axis) versus the predicted values  by the MLP model (y axis) for train and test datasets (maroon and salmon, respectively). C) Acceptance rate along the MCMC steps, D) scale reduction factor $\hat{R}$ of coordination number (circles), radius of gyration (squares), and potential energy (triangles) as a function of the MCMC step and E) as a function of the corresponding wall-clock time, for ab-flowMC (red) and ab-flowMC w.o. MLP (blue). We also included in C), D), E) as black lines the results from an ab-flowMC with $\epsilon_{\rm DFT}=0.3$  simulation that used a MLP trained a-priori with 10000 random samples generated from a 12-dimensional normal distribution and 5000 samples from a MD trajectory. 
}
\label{fig:fig2}
\end{figure}

Using 500 configurations from the short MD simulations as initialization, we applied the ab-flowMC algorithm for three settings: $(i)$ ab-flowMC (with $\epsilon_{\rm DFT} = 1$ for the five first cycles, $\epsilon_{\rm DFT} = 0.5$ for the five next cycles and $\epsilon_{\rm DFT} = 0.3$ for the rest of the simulation), $(ii)$ ab-flowMC without (w.o.) MLP predictions (where all energy evaluations were performed with DFT) and $(iii)$ ab-flowMC with fixed $\epsilon_{\rm DFT}=0.3$  and a a pre-trained MLP with a large dataset of 15000 samples. For $(i)$ and $(iii)$,  we  actively train the MLP every 5 cycles.
In \cref{fig:fig2}, we show the results for the
bipyramidal isomer (similar results for the planar isomer are shown in Supplementary \cref{app:fig:single-isomer-planar}). 

Conformational samples from the initial MD trajectory, and ab-flowMC are shown in \cref{fig:fig2}A as a function of two collective variables, the radius of gyration and the coordination number (see the Supplementary Text). The density of samples from ab-flowMC w.o. MLP is showed as blue dashed lines. 
Because the MD simulation is short, the MD samples may not fully cover the entire basin, for example, in the planar isomer the ab-flowMC samples cover a wider region as shown in Supplementary \cref{app:fig:single-isomer-planar}A. We find a good overlap between the histograms of potential energies from the different algorithms (\cref{fig:fig2}A inset) and a good correlation between the DFT and MLP-predicted energies for the MCMC samples (\cref{fig:fig2}B), showing that the MLP can learn to predict energies accurately with the benefit of computational gain. The adaptive workflow enables the acceptance rate of the flow proposals in the Metropolis-Hastings to increase with the active training (\cref{fig:fig2}C). We find a good correlation between the negative log-likelihood of the flow proposals and potential energies (Supplementary \cref{fig:nlls-ab-flowMC}).   
Calculations of the potential scale reduction factor $\hat{R}$ \cite{vehtariRanknormalizationFoldingLocalization2021} show that the MCMC converges at a similar rate, with or without MLPs, as a function of the number of steps (\cref{fig:fig2}D). Moreover, the results confirm a computational speed up factor close to two-fold 
in wall clock time when using the MLP pre-trained with only 500 steps (red line in \cref{fig:fig2}E). However, the computational gain depends on the complexity of the energy landscape, as exposed in Supplementary \cref{app:fig:single-isomer-planar}E for the planar isomer. Interestingly, we find that pre-training the MLP with a large dataset is less advantageous as it gives similar computational costs as ab-flowMC w.o. MLP (black and blue lines in \cref{fig:fig2}E). The ML losses during active training are given for reference(Supplementary \cref{fig:mlploss} and \cref{fig:nfloss}).

\subsubsection{The optical spectra of Ag$_6$ isomers}

\begin{figure}[h]
\centering
\includegraphics[width=0.45\textwidth]{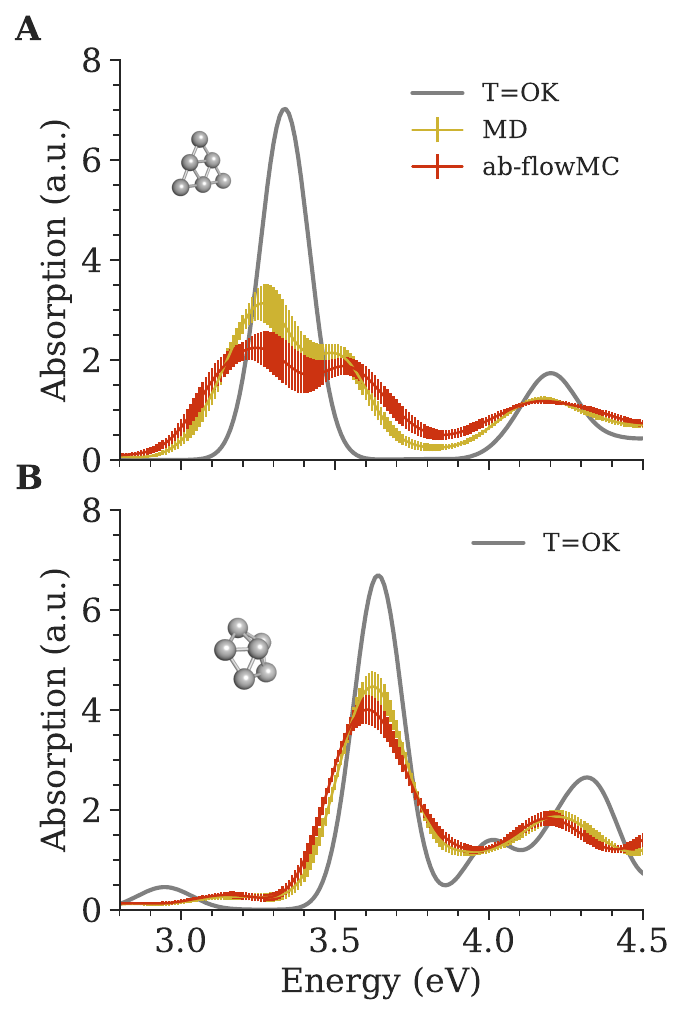}
\caption{Optical spectrum for the Ag$_6$ planar(A) and bipyramidal (B) isomers using the lowest-energy configuration at $T=0 {\rm K}$ (black), MD samples (yellow) and ab-flowMC configurations (red) with a confidence interval of 95\%. 
}
\label{fig:optical-spectrum}
\end{figure}

Since ab-flowMC produces Boltzmann-distributed samples around each isomer, we can estimate local thermodynamical observables. In this section, we will focus on the optical spectrum. 
The measured spectrum of Ag$_6$ embedded in a neon matrix at very low temperature shows mainly one intense and narrow transition at 3.45 eV with a shoulder at 3.65 eV \cite{Lecoultre2011}. This result suggests coexistence of the planar and bipyramidal isomers in the sample. With Time Dependent DFT (TDDFT), we simulate the spectrum of both isomers below 4.0 eV. We find one intense degenerated excitation at 3.34 eV and 3.64 eV respectively (see $T=0 {\rm K}$ curves in \cref{fig:optical-spectrum}). 
Using ab-flowMC samples, we compute the absorption spectrum to access the optical spectrum weighted by its Boltzmann factor around each state (red in \cref{fig:optical-spectrum}). We compared this to the average spectrum using the same number of MD configurations (yellow in \cref{fig:optical-spectrum}). We find important differences between the $T=0 {\rm K}$, MD and MCMC spectra. Isomer 0's main intense peak at $T=0 {\rm K}$ around 3.45 eV breaks into two peaks which can be associated with the breaking of the planar symmetry. The MD spectrum is in agreement with the MCMC double peak spectrum although with different relative intensities. From the MCMC prediction, it appears that an optical spectrum with two peaks in the range 3.0-4.0 eV can be interpreted as a modified non-planar isomer 0 spectrum rather than an isomerization of Ag$_6$.

\subsubsection{Free-energy computations}

\begin{figure}[h]
\centering
\includegraphics[width=\textwidth]{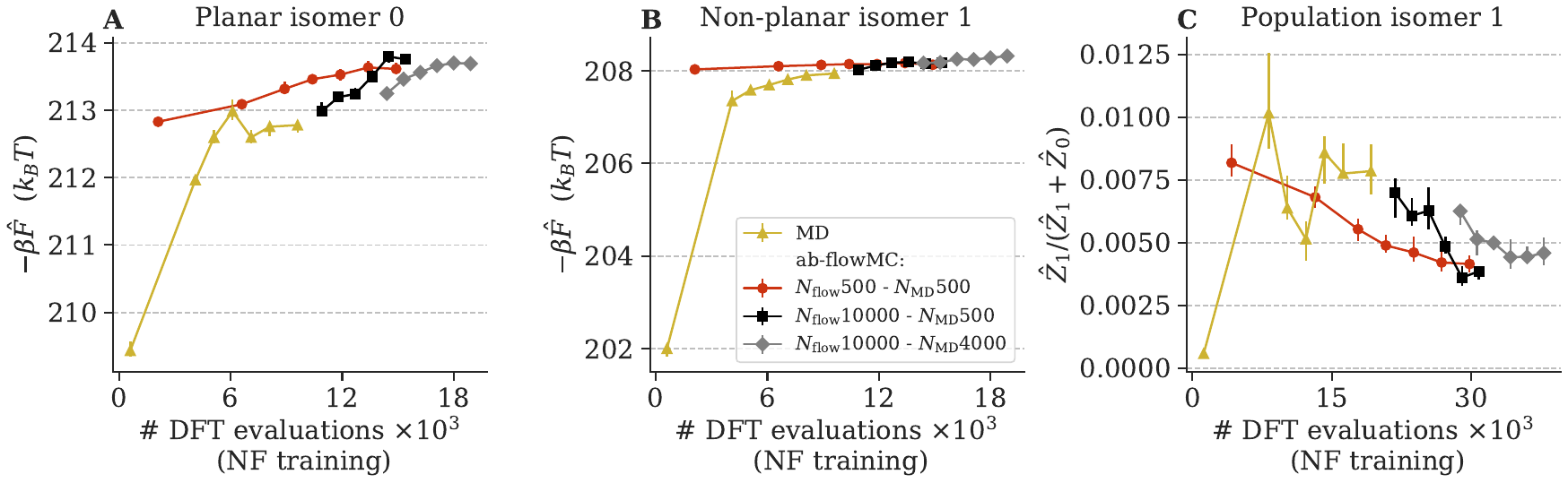}
\caption{BAR absolute free-energies for the planar isomer 0 (A), bipyramidal isomer 1 (B) and relative population prediction (C) computed with a NF trained with pure MD samples (yellow) or with ab-flowMC starting with $N_{\rm MD}$ initial samples from MD and DFT energies evaluated on $N_{\rm flow}$ random samples from the initialized flow (see legend for the details of the combinations tested). For $N_{\rm MD}=500$ and $N_{\rm flow}=500$, $\epsilon_{\rm DFT}$ follows the decreasing schedule from $1$ to $0.3$ (red). For larger initial datasets, $\epsilon_{\rm DFT}=0.3$ is fixed (gray and black). The cost of initial samples is included in the computational budget reported on the x-axis. Error bars represent the standard deviation from repeating the experiment 10 times with independent MCMC and NF samples.}
\label{fig:absolute-fe}
\end{figure}

The flows can be leveraged to compute free-energy differences with TFEP or BAR as previously proposed by several works \cite{wirnsberger_targeted_2020,ding_computing_2020, wirnsbergerNormalizingFlowsAtomic2022}. Here, we focus on BAR as it was found to be better behaved \cite{wirnsberger_targeted_2020}. The absolute free-energy of each state is estimated by applying the BAR framework to compute the free-energy difference between the normalized flow density $\pnf_{0}$ (respectively $\pnf_{1}$) and the Boltzmann distribution density associated with MLP energy $\Umlp_{0}$ (respectively $\Umlp_{1}$) (see the Methods). Importantly, the BAR estimator assumes independent samples from each distribution and requires a large quantity of such samples to mitigate variance. For $\pnf_{0}$ and $\pnf_{1}$, these samples are cheap to obtain by construction. For each isomer state, we run long Metropolis-Hastings MCMCs (as in box 2 of \cref{fig:algo-sketch}) relying exclusively on the MLPs to predict the energies, 
allowing us to obtain $20$ chains of $10,000$ steps in a few minutes.
We then employ the flows and MCMCs samples in the BAR computation, again using exclusively the MLPs to compute the energies. This was repeated with flows extracted from different cycles along the ab-flowMC simulation and the resulting negative free-energy estimates $- \hat F$ are plotted on \cref{fig:absolute-fe} as a function of the number of DFT evaluations necessary for training. We also compare ab-flowMC with a decreasing schedule for $\epsilon_{\rm DFT}$ and with a fixed $\epsilon_{\rm DFT}=0.3$ pre-trained with a larger dataset. We note that relying exclusively on MD simulations, which would correspond to an extension of the DeepBAR method \cite{ding_computing_2020} to \emph{ab initio} accuracy, is not computationally feasible due to the length of the simulation needed to extract independent configurations. For comparison purposes, we instead replicate the experiment using flows trained with \emph{ab initio} MD trajectories exclusively, but retain the rest of the protocol similar as above: the ab-flowMC trained-MLP is used to compute energies of flow proposals.
The predicted absolute free-energies are found to increase with the length of the MD or MCMC trajectories used to train the flow (\cref{fig:absolute-fe}A and B). We hypothesize that this is due to longer chains exploring their free-energy minima basin further and thereby capturing more reliably the entropic contribution. Noticeably, ab-flowMC training (red lines) requires much less DFT computations than MD-based training (yellow lines) to approach a converged value of the free energies, which is consistent with the weaker conformational exploratory power of MD compared to ab-flowMC. As above, we find that pre-training the MLP with a larger dataset is less advantageous and keeping $\epsilon_{\rm DFT}$ fixed is less advantageous (black and gray lines). In the Supplementary, we show how overfitting in a flow model translates into errors in the free-energy prediction (Supplementary \cref{fig:overfitting}). The absolute free-energy estimate of both isomers can be combined to predict the relative population of the non-planar bipyramidal isomer 1 which is given by $\hat Z_1 / (\hat Z_1 + \hat Z_0) $ with $\hat Z_{i} = e^{-\beta \hat{F}_{i}}$ for $i=\{0,1\}$ (\cref{fig:absolute-fe} C). The predicted population fluctuates until convergence of the free energy estimate is reached for each state, which explains the larger fluctuations for the MD-based approach. Overall, the relative population predicted is around $0.5\%$ that is significantly below the predictions of $10\%$ made at $T=300K$ by metadynamics \cite{cramer2004, sucerquia2022}.

\subsubsection{Ab-flowMM MCMC}

\begin{figure}[h]
\centering
\includegraphics[width=\textwidth]{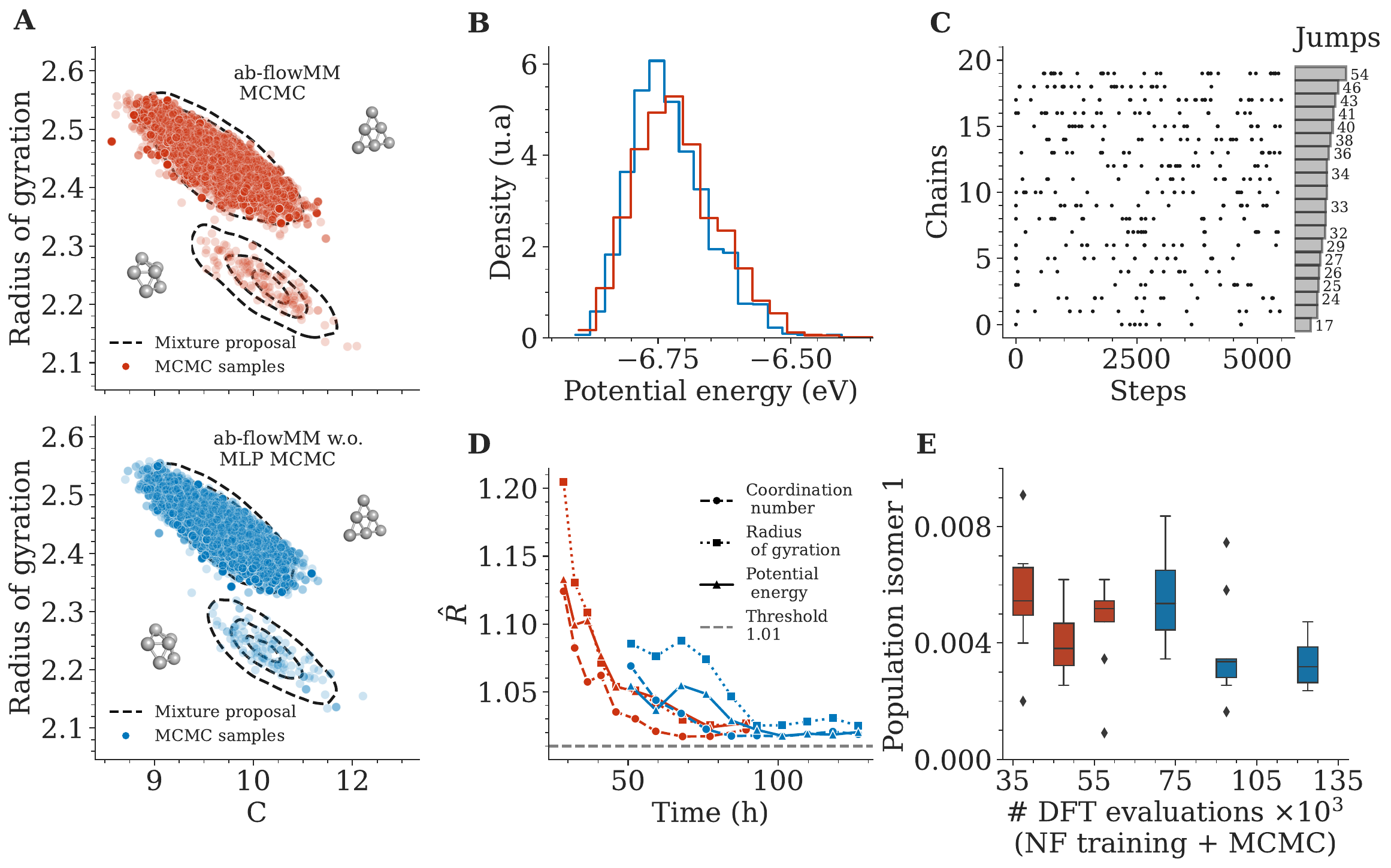}
\caption{MCMC across isomers with the ab-flowMM mixture model built using the NFs and MLPs from ab-flowMC. A) Ab-flowMM proposal (black dash-lines), and MCMC samples with MLP (top, red) and w.o. MLP using 100\% DFT energy evaluations (bottom, blue). B) Potential energy distributions of the samples from the two settings. C) Ab-flowMM MCMC isomer state visited at each step. Visits to the bipyramidal isomer 1 are marked with a black dot, and to isomer 0 are left in white. Chains are sorted by total number of visits to isomer 1, which is reported in the histogram on the right.  D) Scale reduction factor$\hat{R}$ as a function of wall clock time. E) Average population of isomer 1 as a function of the number of DFT evaluations used during flow training and the ab-flowM simulation using flows from different cycles of ab-flowMC. Box plot displays the variability across batches of 20 chains using the last 2750 steps. 
}
\label{fig:fig4}
\end{figure}

We can leverage the flows and MLPs from ab-flowMC to build a mixture model, ab-flowMM, as described in the Methods.  
Using this mixture model as a proposal in a Metropolis-Hastings MCMC (see details in Supplementary Text \ref{app:sec:mixture-mcmc}) enables sampling across the different states without having to wait exponentially long times to cross the barrier as in regular MD. In particular, the relative population of both isomers can be estimated from this MCMC procedure. We compared running the ab-flowMM MCMC using MLPs with retraining (DFT evaluation on 20$\%$ of proposals) and using only DFT for the energy evaluations (ab-flowMM w.o. MLP). In \cref{fig:fig4}A, MCMC samples projected along the two chosen collective variables are distributed similarly for both cases, presenting a wider conformational basin for the planar isomer 0. The sampled energy distributions are also similar (\cref{fig:fig4}B) and transitions between the two metastable states regularly occur for all chains (\cref{fig:fig4}C). For both cases, we find that the acceptance rate is stable as the MCMC progresses (Supplementary \cref{fig:acceptace-ab-flowMM}).  In \cref{fig:fig4}D, we report the $\hat{R}$ factor as a function of the wall-clock time, which approaches the common heuristic threshold of convergence of $1.01$. 
The relative population of the isomers converge to similar values with or without MLPs (\cref{fig:fig4}E), and these values are also consistent with the populations predicted with BAR above. 

\section{Discussion}

We developed an efficient active-learning approach to obtain Boltzmann-distributed molecular configurations and potential energies with quantum mechanical accuracy. For each isomer, ab-flowMC actively trains a NF to propose reliable configurations and an MLP to predict energies using an adaptive MCMC. The main advantage of this methodology is that learning of the surrogate models for the distribution and the energies is done simultaneously by actively improving within the true regions of interest that are indicated by the MCMC. After convergence, the NFs and MLPs enable a fast and accurate application for a wide variety of studies, from optical spectrum calculations to free-energy differences. 
This work shows that it is possible to actively learn both the distribution of configurations and \textit{ab initio} energies using MCMC.    

The ab-flowMC method has great potential but there are several issues that can be addressed in future work. For example, for systems requiring higher quantum precision, more sophisticated MLP architectures \cite{musilPhysicsInspiredStructuralRepresentations2021} than the one used for Ag$_6$, should be used. 
Ab-flowMC is not designed to sample around transition states or improbable conformations, therefore, large uncertainties in the MLP prediction of these rare states should be expected and could hinder the estimation of free-energy barriers.
However, it could be combined with alternative methodologies (\textit{e.g.}, refs. \cite{duan2023accurate,jung2023machine}) to extract the full molecular thermodynamic picture. Moreover, the scaling of the method to systems with thousands of atoms will probably require optimization and further developments \cite{ceriottiInefficiencyReweightedSampling2012}.

\section{Methods}


\subsection{\textit{Ab initio} Molecular Dynamics}

\textit{Ab initio} MD incorporates electronic structure theory into classical MD. The fundamental idea is based on computing the forces of the nuclei on-the-fly with the electronic structure method while the trajectory is generated. In our case, we used Kohn-Sham DFT  \cite{marx_hutter_2009}~. We used the python package ASE (Atomic Simulation Enviroment) \cite{ase-paper} and the DFT library GPAW \cite{Mortensen2005, Enkovaara2010} to run \textit{ab initio} MD and compute the molecules' properties. For the DFT, we used the LCAO-pvalence basis set of atomic orbitals and the exchange correlation functional PBE \cite{Perdew1996}~. The cell size was 16 \AA~and the real-space grid spacing was 0.2 \AA. We used the Andersen thermostat provided by ASE and ran an NVT simulation at T=350K for each isomer. The time step employed was 10 fs and the probability factor was 0.1. %
The molecular conformations were saved every step, however, we dismissed the first 200 steps (0.002 ns) until thermal equilibration was reached. We ran the MD for a total of 700 time steps, yielding initial datasets of 500 MD configurations for training the NFs.  We also ran ab-flowMC using a longer MD trajectory (4000 steps) at initialization. For reference, we report the free-energy estimation results with this more costly initialization in \cref{fig:absolute-fe}. Finally, we ran an MD simulation of 10,000 steps to compare with methods that use only MD configurations.

\subsection{Optical spectra}
To compute the optical spectra, we used linear-response time-dependent DFT (LR-TDDFT) where the Casida equation is solved in the basis of Kohn–Sham (KS) particle–hole transitions in the frequency domain \cite{Gross84, Casida95}. We used the linear-response approach provided by GPAW with a maximum energy difference for Kohn-Sham transitions parameter of 5.5eV \cite{Walter2008, Kuisma2015, Rossi2017}. LR-TDDFT spectra presented in this work are Gaussian-broadened with width $ \sigma= 0.1$ eV.
We compute the optical spectrum at $T=0$K using the configuration from the local energy minimum (obtained from minimization) of each isomer. For the MD optical spectrum, we used 50 samples from a 10,000 steps MD trajectory, taking one sample every 10 steps starting from 9500 step. For the ab-flowMC spectrum, we used the last configuration visited by  each Markov chain, resulting also in 50 independent samples.

\subsection{Internal coordinates and real centered representation}

ASE uses Cartesian coordinates to describe the positions of the atoms of the molecule. However, we used internal coordinates to account for translation and rotation invariance. This representation reduces the dimensionality of the system from 18 to 12 coordinates.
We use the Python library Chemcoord \cite{chemcoord-paper} to compute the internal coordinates from the ASE object atoms. In addition, we centered the data by subtracting the internal coordinate values from the minimum energy structure of each isomer, and we applied the tangent function to the angles to avoid their otherwise bounded support. 
This final representation of real centered coordinates is used for all the inputs of our machine learning models. 
The details of these transformations and induced Jacobians are given in Supplementary Text \ref{app:sec:coordinates}.

For visualization purposes, we make use of two CVs: the coordination number and radius of gyration, which yield a projection where the two isomer basins are separated. These CVs are described in the Supplementary Text \ref{app:sec:cv}.

\subsection{Neural networks' architectures and training losses}


Normalizing flows are a type of generative model relying on a base distribution and a parametrized transformation that is invertible and differentiable to build probability distributions \cite{Papamakarios2021}. The expressiveness of the flow depends on the architecture.
We used the RealNVP architecture \cite{realnvp-paper}, which relies on the composition of affine coupling layers. At layer $\ell$, its input $x^\ell$ is partitioned into two sub-vectors $x^\ell = (x^\ell_1, x^\ell_2)$. Part of the layer outputs is unchanged $x^{\ell+1}_1 = x^{\ell}_1$ and the rest is transformed with an affine function $x^{\ell+1}_2 = s_\theta(x^{\ell}_1) * x^{\ell}_1 + t_\theta(x^{\ell}_1)$, where the component-wise scaling and translation coefficients are neural networks to parametrize. Across layers, the part that is copied and the part that is updated are alternated. 
In our experiments, we use 4 coupling layers, and each scaling and translation network is a fully-connected multi-layer perceptron with 3 hidden-layers each of width 64 units.

We denote by $T_\theta$ the diffeomorphism of $\mathbb{R}^{12}$ defined by the composition of all the coupling layers. The push-forward distribution parametrized by the flow is given by 
\begin{equation}
    \pnf_\theta(x) = p_B(T_\theta^{-1}(x))|\det \nabla T_\theta^{-1}(x)|~, 
\end{equation}
with $\pbase$ the Gaussian base distribution and $\nabla T_\theta^{-1}(x)$ the Jacobian of the inverse transformation. The covariance of the Gaussian base distribution is taken equal to the empirical covariance of the initial MD configurations in $\DNF$.
The parameters $\theta$ are adjusted to minimize the negative log-likelihood that is defined, given a dataset of configurations $\DNF = \{x_i\}_{i=1}^n$, as
\begin{equation}
    {\rm NLL}(\DNF, \theta) = - \sum_{i=0}^{n} \log{\pnf_\theta(x_i)}~.
\end{equation}

For the machine learning potential we use simple multi-layer perceptrons with 3 hidden layers, each with 64 units. We regress the potential energy of each isomer in real centered coordinates minimizing the mean squared error on the $\DMLP$ dataset of configuration-energy pairs $(x, \Edft(x))$.

\subsection{Ab-flowMC hyperparameters}
\label{app:sec:ab-flowMChyper}

Among the 500 configurations extracted from the short MD trajectory, we randomly choose 400 configurations to build the training dataset of the NF, $\DNF$. We randomly select half of the remaining 100 samples to initialize 50 Markov chains. 
At each cycle of the ab-flowMC workflow (\cref{fig:algo-sketch}), all chains are prolonged for 20 steps. We add the latest MCMC configurations to the $\DNF$ dataset and use them to retrain the flow at each cycle. 
The learning rate is 10$^{-4}$ and 400 steps of full-batch gradient descent are used at each cycle. 

The training dataset for the MLP, $\DMLP$, consists of 500 configurations from the short MD trajectory and 500 random samples from the initialized NF. 
We initialize the parameter $\epsilon_{\rm DFT}=1$ such that we compute the energy of all configurations with DFT for the first 5 cycles. Then, we decrease $\epsilon_{\rm DFT}$ to $0.5$ and select randomly at each cycle 50 \% of proposals for DFT energy evaluation for another 5 cycles. The parametrer $\epsilon_{\rm DFT}$ is kept fixed to $0.3$ for the rest of the simulation.  All configurations and energies that are evaluated with DFT are added to $\DMLP$, thus, increasing the dataset by $7500$ during the first 10 cycles and by $1500$ samples every 5 cycles thereafter. 
A random split of the train and test data is operated at each data addition. The MLP is retrained every 5 cycles using a learning rate of $5\times10^{-5}$, 30000 steps of gradient descent, and a batch size of 1000. 


\subsection{Metropolis-Hastings MCMC}
\subsubsection{Accept-reject step}
We now explain the accept-reject step of the Metropolis-Hasting algorithm in ab-flowMC (box 2 in \cref{fig:algo-sketch}). Starting from a previous position $x_t$ with energy $U_t$, we consider the proposed new configuration $\tilde x_{t+1}$ with energy $\tilde U_{t+1}$. The next position of the Markov chain is generated as $x_{t+1}, U_{t+1} = \Gamma_{\rm MH}(x_{t}, U_{t}, \tilde x_{t+1}, \tilde U_{t+1})$ with
\begin{equation}
    \Gamma_{\rm MH}(x_{t}, U_{t}, \tilde x_{t+1}, \tilde U_{t+1}) = 
    \left\{ 
    \begin{array}{ll}
         \tilde x_{t+1}, \, \tilde U_{t+1} & \text{w. prob. }  \, \min \left( 1,  
         \frac{\pnf_{\theta}(\tilde x_{t+1}) e^{-\beta U_t}}{\pnf_{\theta}(x_{t})e^{-\beta \tilde U_{t+1}} }  \right) \\
         x_t ,\, U_t& \text{otherwise}
    \end{array}
    \right. ~,
\end{equation}
where $\pnf_\theta(\cdot)$ is the NF's probability density and $\beta = 1/(k_{\rm B}T)$.

The corresponding procedure when using the mixture proposal ab-flowMM is described in Supplementary Text \ref{app:sec:mixture-mcmc}.

\subsubsection{$\hat{R}$ convergence test for Markov chains}
The potential scale reduction factor $\hat{R}$ is a common convergence test for parallel MCMC chains that can be roughly interpreted as comparing the variance within chains with the variance across chains. We used the implementation of the open source python package ArviZ \cite{kumarArviZUnifiedLibrary2019} which implements a more robust rank-normalized version of  $\hat{R}$ \cite{vehtariRanknormalizationFoldingLocalization2021}~. 
By construction $\hat{R} \geq 1$ in all cases and $\hat{R}=1$ at convergence. We use the commonly considered threshold of $\hat{R}=1.01$ to assess convergence. 

\subsection{BAR calculations}
The two NFs, from ab-flowMC for each state, provide an approximation of the Boltzmann distribution of the conformers. They can be used to compute the relative free energy between both states using Targeted free energy perturbation (TFP) \cite{jarzynski_targeted_2002,wirnsberger_targeted_2020} or the Bennett acceptance ratio (BAR) \cite{Bennett1976,jia_normalizing_2019,ding_computing_2020}. The idea is to compute the free-energy difference between the NF modeling one of the conformers and the Boltzmann distribution restricted to the conformer's basin. We chose to focus on the BAR method, as it is typically better behaved \cite{wirnsberger_targeted_2020}.

The BAR method estimates ratios of partition functions. We define the partition function of each isomer state $(i)$ as
\begin{align}
    {\Zdft^{(i)}} = \int_{\{ x_1, x_2, \R^{12}\}}{e^{-\beta \Eneutral(x)}\mathbb{1}_{x \in \mathcal{D}^{(i)}}{\rm d}x }~,
\end{align}
where $\Eneutral(x)$ is the potential energy, $\{ x_1, x_2, \R^{12}\}$ means that we fix the position of the two first atoms to fix the a global translation and rotation and $\mathbb{1}_{x \in \mathcal{D}^{(i)}}$ is the indicator function that $x$ is in the domain of isomer state $(i)$. In practice, one need not explicitly define isomer states $\mathcal{D}^{(i)}$ in our implementation -- provided that the states are well separated in the configurational space -- since all expectations will be approximated by a Monte Carlo empirical average relying on samples for which there is no ambiguity of the state they belong to. The partition function associated with the flow state can be formally defined from $\pnf_{\theta,i}(x) = p_B(T_{\theta,i}^{-1}(x))|\det \nabla T_{\theta,i}^{-1}(x)|$ with $\pbase(y) = e^{-\Ebase(y)}/\Zbase$, such that $\pnf_{\theta,i}(x) = e^{-U_{\theta,i}(x)} / \Zbase$ with
$U_{\theta,i}(x) = \Ebase({T_{\theta,i}}^{-1}(x)) - \log |\det \nabla {T_{\theta,i}}^{-1}(x)|$. In practice, the base partition function $\Zbase$ is known, typically chosen equal to 1.

We denote by $\langle\cdot\rangle_f$ the expectation with respect to a probability density proportional to the positive function $f$. The BAR derivation starts from the observation that for any scalar function $w(\cdot)$ defined on the coordinate space
\begin{align}
   \frac{\Zdft^{(i)}}{\Zbase} = \frac{
   \left\langle w(y) e^{- \beta \Eneutral(T_{\theta,i}(y))} \right\rangle_{\pbase}
   }{
   \left\langle w({T_{\theta,i}}^{-1}(x)) e^{- \Ebase({T_{\theta,i}}^{-1}(x)) - \log |\det \nabla T_{\theta,i}^{-1}(x)|} \right\rangle
   _{e^{-\beta \Eneutral}\mathbb{1}_{\mathcal{D}^{(i)}}}
   } 
\end{align}
where in the right-hand side the numerator is an expectation over $y\sim \pbase(y)$ and the denominator is an expectation over $x\sim e^{-\beta \Eneutral(x)}\mathbb{1}_{x \in \mathcal{D}^{(i)}} / {\Zdft^{(i)}}$ --the Boltzmann distribution restricted to the isomer state of interest--. This equality leads to an estimator of free-energy differences based on samples from the latter two distributions. Choosing the function $w(\cdot)$ to optimize the variance leads to the estimator $\hat{r}^{(i)} = {\widehat{\Zdft^{(i)}}}/{\Zbase}$ where $\hat{r}^{(i)}$ is solution of the equation 
\begin{align}
    \sum_{k=1}^{n_*} \frac{\hat{r}^{(i)} n_\theta  e^{-U_{\theta,i}(x_{*,k})}}{n_* e^{-\beta \Eneutral(x_{*,k})} + n_\theta \hat{r}^{(i)} e^{-U_{\theta,i}(x_{*,k})} } = \sum_{k=1}^{n_\theta} \frac{n_* e^{-\beta \Eneutral(x^{(i)}_{\theta,k})}}{n_* e^{-\beta \Eneutral(x^{(i)}_{\theta,k})} + n_\theta \hat{r}^{(i)} e^{-U_{\theta,i}(x^{(i)}_{\theta,k})} }. 
\end{align}
where $\{x^{(i)}_{\theta,k}\}_{k=1}^{n_\theta}$ are samples from the flow and  $\{x_{*,k}\}_{k=1}^{n_*}$ are approximate samples from the Boltzmann distribution obtained by a Metropolis-Hastings MCMC enhanced by the same flow (same as box 2 of \cref{fig:algo-sketch}).
Our implementation was inspired by the open source package pocoMC \cite{karamanis2022pocomc}.

\subsection{\emph{Ab initio} flow Mixture Model and MCMC}

We can combine the NFs and MLPs that were trained separately for each state using ab-flowMC to create a mixture model, referred to as ab-flowMM. Let $\pnf_{\theta, 0}$ and $\pnf_{\theta, 1}$, be the NFs learned densities for the planar isomer 0 and bipyramidal isomer 1, respectively. The NFs can be combined to make a ``full'' density mixture model $\pnf_{\mathrm{MM}}=w_0\pnf_{\theta, 0}+w_1\pnf_{\theta, 1}$, where the real positive numbers $w_0$ and $w_1$ are the relative weights of each flow with $w_0+w_1=1$. To obtain consistent statistics across basins, we then run a Metropolis-Hastings MCMC, similar to the box 2 in \cref{fig:algo-sketch}, but using the ab-flowMM $\pnf_{\mathrm{MM}}$ as a proposal.
Let $\Umlp_{\alpha, 0}$ or $\Umlp_{\alpha, 1}$ be the corresponding trained MLPs. Using an extra state variable tracking the isomer state of each $\tilde{x}_t$, we can select the appropriate MLP for the accept-reject step at each iteration, when not using DFT, and pursue the active learning of the MLP on proposals selected for DFT evaluation.  Further details are described in Supplementary Text \ref{app:sec:mixture-mcmc}.

In the presented experiment, we ran 20 MCMC chains for 5500 steps for ab-flowMM and 3000 for ab-flowMM w.o. MLP. The chains are initialized with MD samples from both isomers in equal proportion, 
and the mixture weights are estimated along the first iterations of the MCMC. For each isomer, weights are initialized at  $0.5$ and updated using $w_i = \alpha * w_i + (1-\alpha) * P_i$ for $i \in \{0, 1\}$, with $\alpha=0.5$ and $P_i$ the relative population of isomer $i$ during the last 10 MCMC steps. We defined maximum and minimum weight values of 0.75 and 0.25, respectively - which are quickly saturated for the Ag$_6$ example with $w_0 = 0.75$ and $w_1=0.25$. 
In comparison to ab-flowMC on a single isomer, we decreased the number of random DFT evaluations to 20\%, $\epsilon_{\rm DFT} = 0.2$. 
For reference, we also ran the experiment with an ab-flowMM proposal built with NFs trained with ab-flowMC without MLP, and propagating the MCMC also without the MLP predictor (\cref{fig:fig4}).

\subsection{Computational resources}
The DFT computations and ML models were run on the Flatiron Institute cluster using 128-core AMD Rome nodes with 1 TB of RAM. 

\section*{Supporting Information}

The Supporting Information includes supplementary texts discussing how to monitor overfitting, details on the changes of coordinates employed, the definition of the collective variables and further details about the ab-flowMM MCMC procedure. Supplementary Figures present results for Ab-flowMC and Ab-flowMM simulations as well as losses and overfitting results. The code base to reproduce the results is presented as Supplementary Information.

\section*{Acknowledgment}
The authors thank Luke Evans, Miguel Caro, and Michele Ceriotti for useful discussions and Christoph Schönle for careful reading of the manuscript. The Simons Foundation has supported this work. M.G. acknowledges funding from Hi! Paris. A. M-T and O. L-A acknowledge funding from the University of Antioquia.  

\bibliographystyle{unsrt}
\bibliography{refs}

\pagebreak

\appendix

\section{Supplementary Text}

\subsection{Overfitting in the NFs training}
\label{app:sec:overfitting} 
The training data of the normalizing flows, extracted from Markov chains, is not stationary before the chains have converged to the limiting Boltzmann distribution. This fact poses a challenge to the monitoring of overfitting, we used the following procedure. At cycle $k$ of the algorithm, we converged independent MCMC chains using the current NF as a proposal and evaluating the accept-reject probability using the best MLP from the entire simulation. Using the MLP allows to run long chains (20 chains of 10000 steps within minutes) to produce converged samples used to form a test set that is used to compute the test log-likelihood. In \cref{fig:overfitting}A and B top, we report the negative test log-likelihood (${\rm NLL_{test}}$) for the ab-flowMC simulations of isomer 0 and isomer 1 for different procedures: $(i)$ where the initial $\DNF$ dataset contained 500 MD conformations and 500 initial flow conformations, with a decreasing $\epsilon_{\rm DFT}$ schedule (red), $(ii)$ and $(ii)$ where the initial $\DNF$ dataset contained respectively $500$ (gray) and $4000$ MD conformations (gray), and for both $10,000$ initial flow conformations with $\epsilon_{\rm DFT}=0.3$ fixed. When the negative test log-likelihood is stable throughout the simulation, we observe that the absolute free energy typically increases and then stabilizes (\textit{e.g.}, red lines in \cref{fig:overfitting}C). However, we observe that in some cases, ${\rm NLL_{test}}$ starts to increase at some point of the simulation, flagging the onset of an overfitting behavior (after which lines are dashed). We observe that overfitting hinders the stability of the absolute free-energy computation (\cref{fig:overfitting}C and D). These errors in the free-energy estimation  logically propagate to isomer population estimation (\cref{fig:overfitting} E). The results highlight the necessity to monitor for overfitting when applying ab-flowMC.

\subsection{Changes of coordinates}
\label{app:sec:coordinates} 
We use the notation $\xyz \in \R^{3N}$ for Cartesian coordinates and $\ic \in \R^{3(N-2)}$ for internal coordinates. We denote by $\FtoIC$ the invertible transformation that maps Cartesian to internal coordinates, augmented with a global translation $\xyz_0 \in \R^{3N}$ and rotation $\phi_0 \in [0,2\pi[^3$
\begin{align}
    \FtoIC(\xyz) = (\xyz_0, \phi_0, \ic)~.
\end{align}
The potential energy $\Eneutral_{\rm cc}: \R^{3N} \to \R$, computed with DFT or predicted by MLP, gives us access to the density of the Gibbs-Boltzmann distribution with respect to the Cartesian coordinates (abbreviated cc):
\begin{align}
    \rho^*_{\rm cc}(x) = \frac{e^{-\beta \Eneutral_{\rm cc}(\xyz)}}{Z_{\rm cc}}~.
\end{align}
To obtain the density on the internal coordinates (abbreviated ic) we need to take into account the Jacobian of the change of variable $\nabla\FtoIC^{-1}(\cdot) \in \R^{3N \times 3N}$, namely
\begin{align}
    \rho^*_{\rm ic}(\xyz_0, \phi_0, \ic) =  \rho^*_{\rm cc}(\FtoIC^{-1}(x_0, \phi_0, \ic))|\det(\nabla\FtoIC^{-1}(x_0, \phi_0, \ic))|~.
\end{align}
We note that the density is invariant with respect to the global translation $\xyz_0$ and rotation $\phi_0$. As a result, not taking them into account only modifies the density's normalization constant and the density in internal coordinates can be written as
\begin{align}
    \rho^*_{\rm ic}(\ic) = \frac{e^{-\beta \Eneutral_{\rm ic}(y)}}{Z_{\rm ic}}~,
\end{align}
with
\begin{align}
    \Eneutral_{\rm ic}(y) = \Eneutral_{\rm cc}(\FtoIC^{-1}(0_{\R^3}, 0_{\R^3}, \ic)) - \frac{1}{\beta} \log|\det(\nabla\FtoIC^{-1}(0_{\R^3}, 0_{\R^3}, \ic))|~,
\end{align}
where by convention the global rotation angles and translation coordinates have been taken equal to the zero vector and angle (both denoted by $0_{\R^3}$).
In addition, we center the representation at the local minimum of each isomer in internal coordinates, $\rc^{i} = \ic - \bar{\ic}^{i}$, where $\bar{\ic}^i$ is the $i$-th isomer local minimum. Let $\rc^{i}_{\angle}$ denote the sub-vector of coordinates of $\rc^{i}$ representing angles. We finally apply component-wise $\rc^{i}_{\angle} \leftarrow \tan(\rc^{i}_{\angle})$.
The tangent transformation adds an additional Jacobian term to get to the final density \emph{real centered} (rc), which is targeted by the machine learning models,
\begin{align}
    \rho^{*,i}_{\rm rc}(\rc) = \rho^*_{\rm ic}(\ic) |\det\left(\nabla \tan(\rc_{\angle})\right)|~.
\end{align}

For simplicity, we use only the notation $x$ in the main text and $\Eneutral$ without the cc or ic, but the changes of variables are carefully tracked according to the above explanation in the implementation. 

\subsection{Collective variables}
\label{app:sec:cv} 
For visualization purposes, we make use of two CVs yielding a projection where the two isomers' basins are separated.
The coordination number $C$ is associated with the number of bonds within the molecule and is defined as 
\begin{equation}
        C = \sum_{i=1}^{N_a} \sum_{j\neq i} \frac{1 - {\left( \frac{r_{ij}}{d} \right) }^6}{1 - {\left(\frac{r_{ij}}{d}\right)}^{12}} ~,
\end{equation}
where d = 2.8 \AA. The radius of gyration $R$ measures the variance of the atomic positions with respect to the center of mass,
\begin{equation}
    R = {\left(\frac{1}{6}\sum_{i=1}^{6} |r_i - r_{CM}|^2\right)}^{1/2} \,~,
\end{equation}
where $r_i$ is the position of the $i$ atom, and $r_{CM}$ is the position of the center of mass.
These CVs were selected from a previous metadynamics study of silver cluster Ag$_6$ \cite{sucerquia2022}~.

\subsection{Details of the ab-flowMM MCMC procedure}
\label{app:sec:mixture-mcmc}

We make use of the mixture model $\pnf_{\mathrm{MM}}=w_0\pnf_{\theta, 0}+w_1\pnf_{\theta, 1}$ as a proposal in the Metropolis-Hastings algorithm. To describe the MCMC update, we introduce a new isomer state tracking variable $\state$ that is either 0 or 1.
At iteration $t$, the Markov chain is at location $x_t$ in isomer state $a_t$ with energy $U_t$. A new state $\tilde a_{t+1}$ is proposed following the probabilities $(w_0, w_1)$. Then, a new conformation $\tilde x_{t+1}$ is sampled from the corresponding NF probability density $\pnf_{\theta,\tilde a_{t+1}}$. The proposal's energy $\tilde U_{t+1}$ is evaluated either using DFT or using the MLP $\Umlp_{\alpha, \tilde a_{t+1}}$. The Metropolis-Hastings is propagated with 
\begin{equation}
    \Gamma_{\rm MH}(x_{t}, U_{t}, \tilde x_{t+1}, \tilde U_{t+1}) = 
    \left\{ 
    \begin{array}{ll}
         \tilde x_{t+1}, \, \tilde U_{t+1} & \text{w. prob. }  \, \min \left( 1,  
         \frac{\pnf_{\rm MM}(\tilde x_{t+1}) e^{-\beta U_t}}{\pnf_{\rm MM}(x_{t})e^{-\beta \tilde U_{t+1}} }  \right) \\
         x_t ,\, U_t& \text{otherwise}
    \end{array}
    \right. ~,
\end{equation} 
where $\beta = 1/(k_{\rm B}T)$.

\clearpage

\setcounter{figure}{0}
\makeatletter 
\renewcommand{\thefigure}{S\@arabic\c@figure}
\makeatother
\section{Supplementary Figures}

\begin{figure}[h]
\centering
\includegraphics[width=\textwidth]{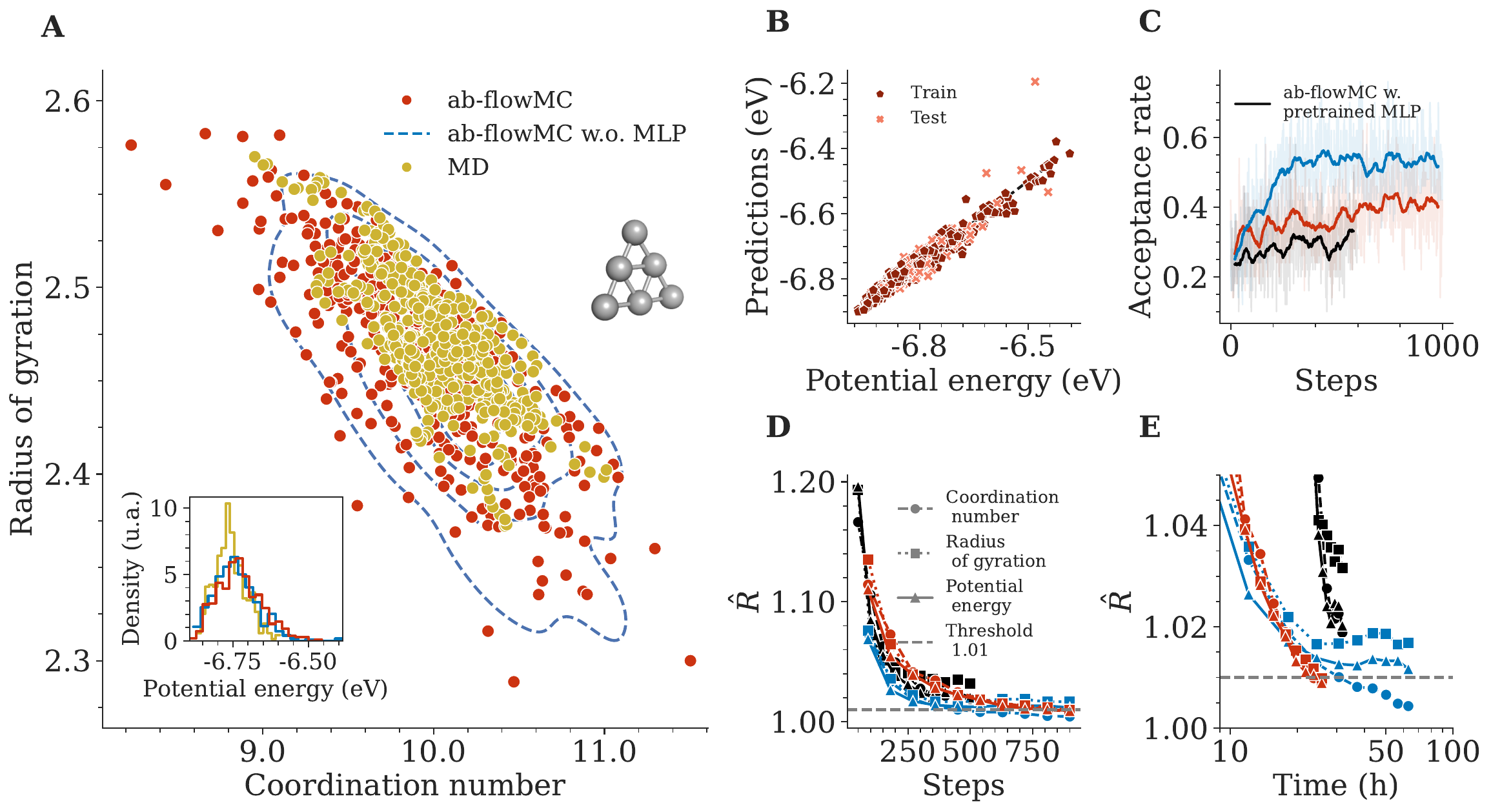}
\caption{
\textit{Ab initio} flowMC for the $Ag_6$ bipyramidal isomer 1. A) Projection along the coordination number and radius of gyration of samples from the MD trajectory (yellow) and the ab-flowMC simulation (red). The density of samples obtained from the ab-flowMC w.o. MLP simulation is displayed for reference (blue dashed lines). Inset: Histograms of the potential energies for all methods. B) DFT potential energy (x axis) versus the predicted values  by the MLP model (y axis) for train and test datasets (maroon and salmon, respectively). C) Acceptance rate along the MCMC steps, D) scale reduction factor $\hat{R}$ of coordination number (circles), radius of gyration (squares), and potential energy (triangles) as a function of the MCMC step and E) as a function of the corresponding wall-clock time, for ab-flowMC (red) and ab-flowMC w.o. MLP (blue).  
}
\label{app:fig:single-isomer-planar}
\end{figure}

\begin{figure}[h]
    \centering\includegraphics[width=0.65\textwidth]{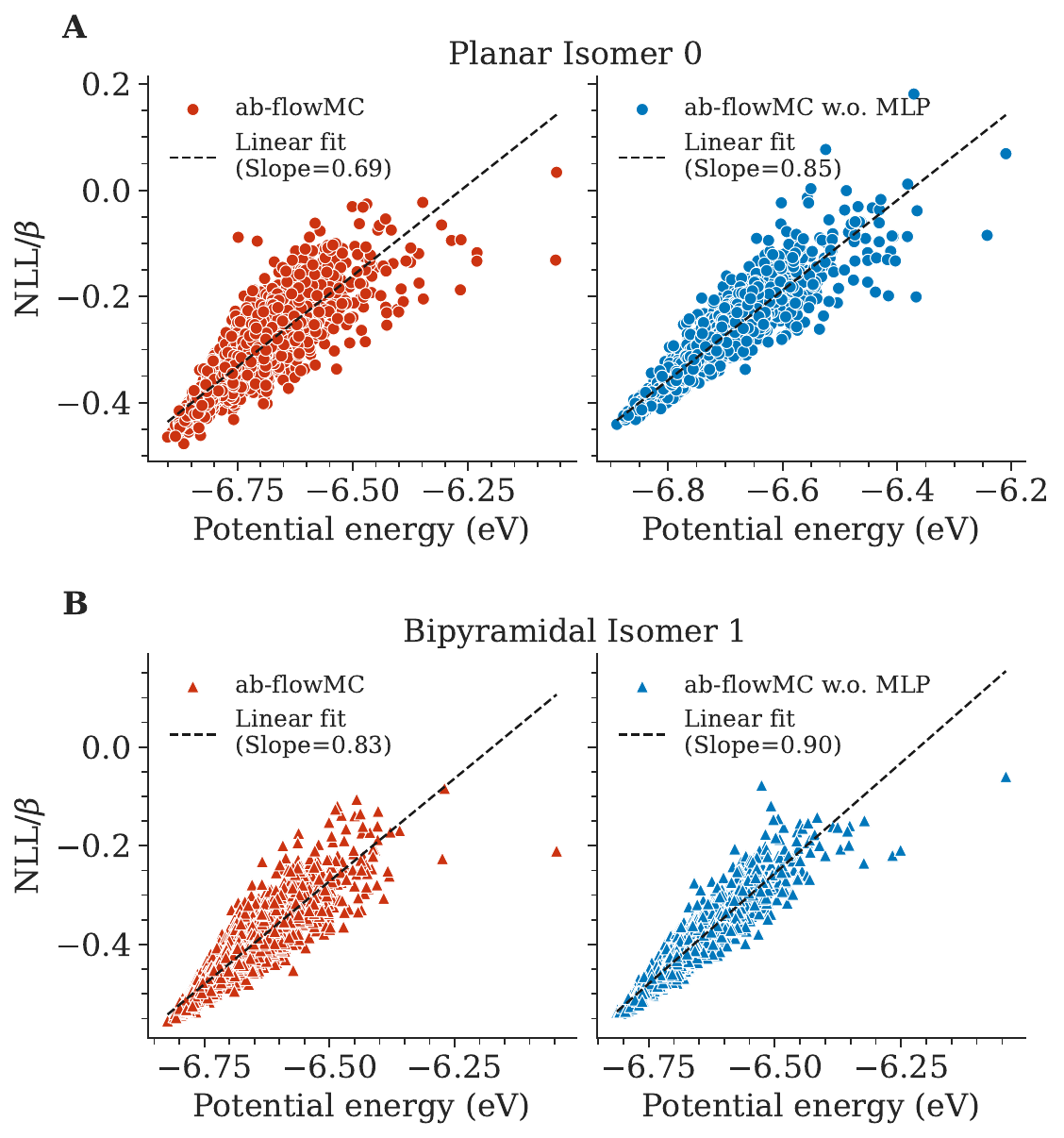}
    \caption{Negative log likelihood divided by $\beta = k_BT$ computed for NF proposed configurations in the ab-flowMC and ab-flowMC w.o. MLP simulations for each isomer at $T=350$K.
    }
    \label{fig:nlls-ab-flowMC}
\end{figure}

\begin{figure}[h]
    \centering
    \includegraphics[width=0.65\textwidth]{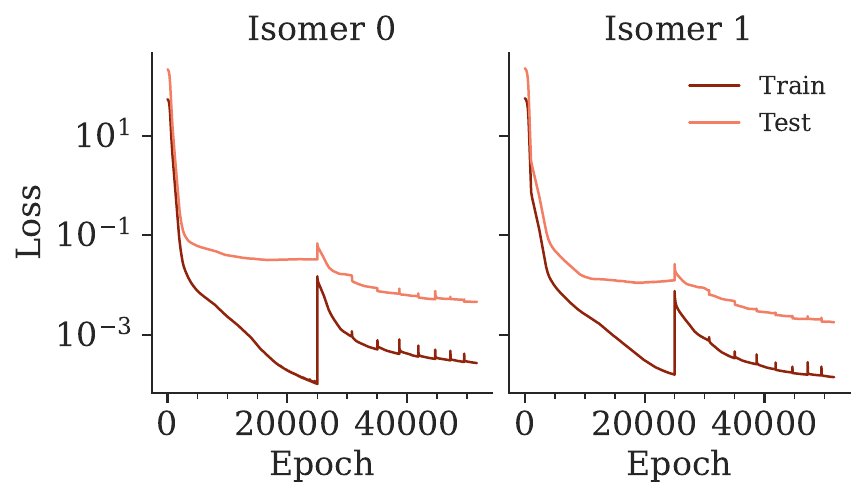}
    \label{fig:ab-flowMM acceptance}
    \caption{Train and test mean squared error while training the MLP models in ab-flowMC for each Ag$_6$ isomer.}
      \label{fig:mlploss}
\end{figure}

\begin{figure}[h]
    \centering
    \includegraphics[width=0.65\textwidth]{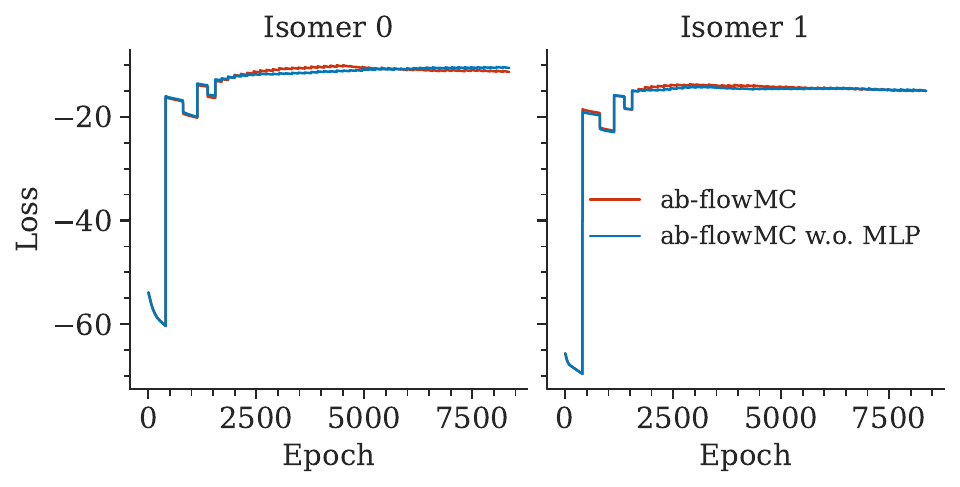}
    \caption{Train negative log-likelihood (NLL) from the adaptive training of the NF models in the ab-flowMC (red) and in the ab-flowMC w.o. MLP simulations (blue) for each isomer.}
    \label{fig:nfloss}
\end{figure}

\begin{figure}[h]
\centering
\includegraphics[width=\textwidth]{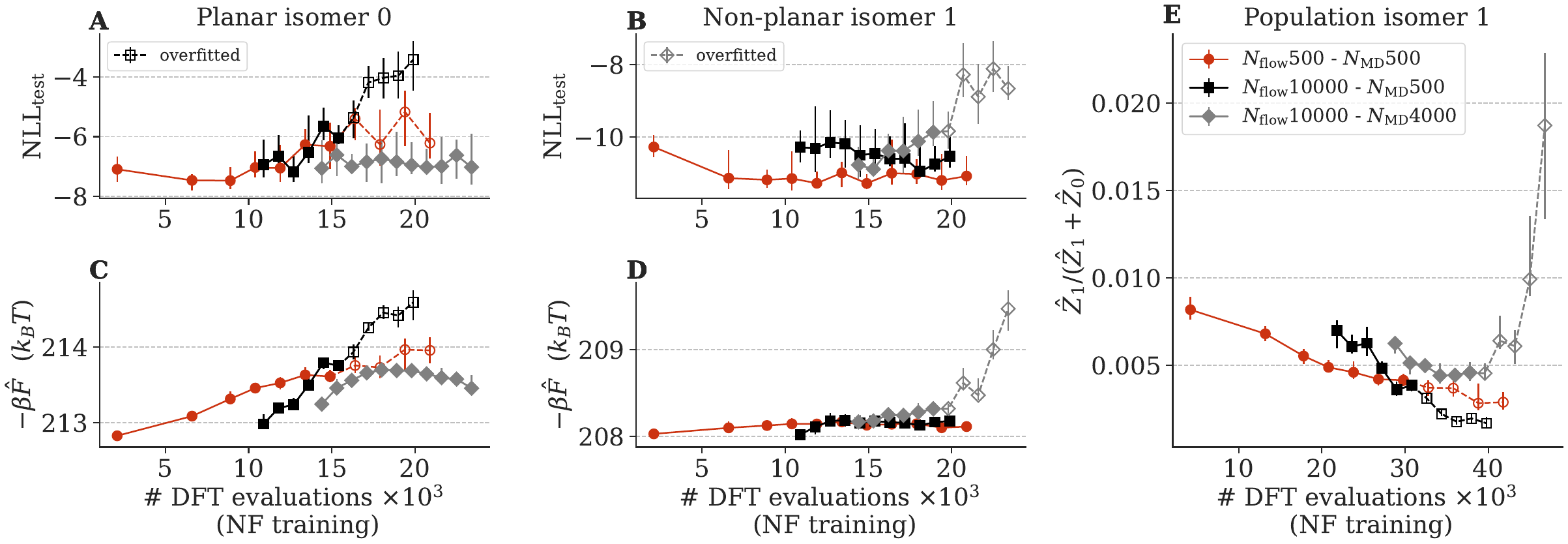}
\caption{Overfitting detection. Results reported in all panels are relative to ab-flowMC simulations started with $N_{\rm MD}$ initial samples from MD and additional DFT energies evaluated on $N_{\rm flow}$ random samples from the initialized flow (see legend for the details of the combinations tested). The cost of these initial samples is included in the computational budget reported on the x-axis. Test negative log-likelihood (${\rm NLL_{test}}$) of the NFs for isomer 0 (A) and isomer 1 (B) at different cycles of ab-flowMC converted to the number of DFT evaluations. (C and D) BAR estimate of the absolute free-energy using the corresponding NFs from (A and B). (E) Relative importance of bipyramidal isomer 1 as estimated using the BAR free-energy estimates of (C and D). Solid lines are turned to dashed when overfitting is detected following a surge of the test negative log-likelihood. The error bars display the standard deviation of the values obtained by repeating 10 times the MCMC generating data for ${\rm NLL_{test}}$ and BAR.}
\label{fig:overfitting}
\end{figure}


\begin{figure}[h]
    \centering
    \includegraphics[width=0.55\textwidth]{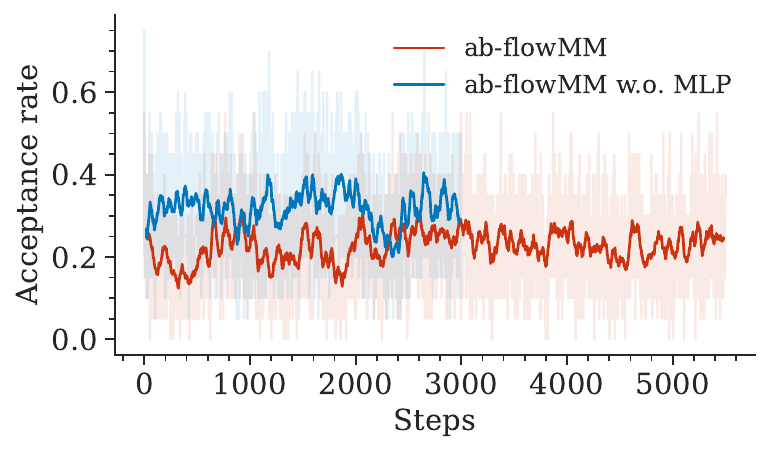}
    \caption{Acceptance rate obtained from running MCMC with ab-flowMM (red) and ab-flowMM w.o. MLP (blue).}
    \label{fig:acceptace-ab-flowMM}
\end{figure}

\end{document}